\documentclass[aps,prd,groupedaddress,11pt,nofootinbib,preprintnumbers]{revtex4-1}

\usepackage{dsfont}
\usepackage{natbib}
\usepackage{amssymb}
\usepackage{graphicx}
\usepackage{amsmath}
\usepackage{url}
\usepackage{epstopdf}
\usepackage{color}
\usepackage{enumerate}
\usepackage[colorlinks=true,breaklinks]{hyperref}

\renewcommand{\H}{\mathcal{H}}
\newcommand{\PP}{P_{\phi}}

\newcommand{\tPP}{\widetilde{P_{\phi}}}
\newcommand{\tm}{\widetilde{m}}
\newcommand{\tlambda}{\widetilde{\lambda}}
\newcommand{\tphi}{\widetilde{\phi}}

\def\be  {\begin{equation}}                                                                 
\def\ee  {\end{equation}}                                 
\def\bea {\begin{eqnarray}}                                                   
\def\eea {\end{eqnarray}}

\begin{document}

\title{Gravity, time and varying constants}

\bigskip

\author{Syed Moeez Hassan}
\email{shassan@unb.ca}
\affiliation{Department of Mathematics and Statistics, University of New Brunswick, Fredericton, NB, Canada E3B 5A3} 

\author{Viqar Husain}
\email{vhusain@unb.ca}
\affiliation{Department of Mathematics and Statistics, University of New Brunswick, Fredericton, NB, Canada E3B 5A3}

\author{Babar Qureshi}
\email{qureshib@mit.edu}
\affiliation{Center for Theoretical Physics, Massachusetts Institute of Technology, 77 Massachusetts Avenue, Cambridge, MA 02139, USA}

\date{\today}

\begin{abstract}

\vskip 1cm
There are  theories which implement the  idea that the  constants of nature may be ``time dependent."   These introduce new fields representing  ``evolving constants," in addition to physical fields. We argue that dynamical matter coupling constants can arise naturally in non-perturbative matter-gravity theories, after a choice of global time is made.  We illustrate the idea in scalar field cosmology with spatial volume as a global clock, and compute the time dependence of the scalar mass and self-interaction coupling constants.   

\vskip 1cm

\end{abstract}

\preprint{MIT-CTP/5037}

\maketitle

 \vskip 0.5cm

\section{Introduction}

Dirac was among the first physicists to suggest that the fundamental constants of nature may be time varying  \cite{Dirac:1937ti}. Since then this suggestion has been widely explored theoretically, and observational constraints from astrophysics and cosmology on various constants have been calculated  \cite{Uzan2011}. Theoretical implementation of this idea  has either been through explicit ad-hoc introduction of time dependent constants, or extra fields in the theory whose dynamics determines the value of constants at any epoch in the Universe's history. Some of these implementations allow both time and space variations of constants. Most work has focused on theories of varying fine structure constant \cite{Bekenstein:1982eu,Barrow:2005hw}, with application to cosmology.

A deeper understanding of this question would involve addressing questions such as what is the ``time" with which constants evolve, which of the fundamental constants are involved, and how a detailed scenario might emerge from a more fundamental theory.  Such questions place the idea squarely in the realm of quantum gravity.  

In this letter we explore the idea of time varying constants by partly exploring these questions. We show how matter coupling constants might become naturally time dependent through the selection of a global time  in a dynamical theory of gravity and matter. This comes about   without introducing special fields. We demonstrate the idea in FRW cosmology using a massive scalar field with quartic self interaction coupled to Einstein's gravity with a cosmological constant.    

That such an  approach is in principle possible follows from considering the Hamiltonian formulation of any theory with general covariance. The Lagrangian theory is background independent in the elementary sense that  the metric is a dynamical field to be varied in the action; in the general case there are no fixed kinematical symmetries, i.e. no Killing vector fields. In particular there is  no global time-like Killing vector field to provide an unambiguous and universal notion of particle and energy. This of course holds in an FRW background where any choice of time function is possible, even in the preferred frame adapted to the spatial Euclidean isotropy group.

 This is not a serious issue for classical general relativity, where measurements are referred to observers with prescribed 4-velocities. For  quantum field theory (QFT), including the graviton, on a curved spacetime, if there is no time-like Killing vector field, then there is no unique   notion of vacuum state. Nevertheless it is possible to compare QFTs  formulated with different time (mode) functions by asking how a vacuum state defined with one notion of time is expanded in  the basis natural for another notion of time. These are the Bogoulubov transformations which play a fundamental role in the derivation of Hawking radiation.
 
 In quantum gravity this problem is further compounded by the fact that there is not even a fixed metric. Instead the full metric is itself subject to quantum fluctuations.  If gravitational dynamics is considered from a fully non-perturbative perspective, which for our purpose means that gravity is not  thought of as the dynamics of small perturbations on Minkowski or any other fixed background spacetime, then there is the famous ``problem of time."  This has many detailed elaborations  \cite{Isham:1992ms,Anderson:2017jij}, but has its ultimate origin in the fact that the Hamiltonian is constrained to vanish, and generates time-reparametrizations rather than ``true" evolution.  

There is no generally accepted solution of this problem, and indeed no clear idea of what a solution might even look like. How then does one pose the question of ``evolving constants" at this level?  From a practical point of view it is possible to simply  choose a function of the phase space variables as time. Examples of this are ``scalar field" time, ``volume time," ``Hubble time," and York time.   The physical Hamiltonian in each of these cases is very different, and the corresponding quantum theories, if they can be constructed at all, would give different physical predictions. 

We will follow the path of selecting  a clock function in the phase space, and study the corresponding physical Hamiltonian. It is known that such Hamiltonians are generically time dependent.  We will see that this time dependence can be absorbed in matter coupling constants. 
 The three fundamental constants  $\hbar, c$ and $G$ do not change in the framework we describe. Instead they serve to define the scale used to measure all other physical constants (henceforth, we will work in $\hbar = c = 8 \pi G = 1$ units).  This would still permit varying fine structure constant through the matter couplings.  
 
\section{Cosmological model} 

The model we consider is gravity coupled to a scalar field in the  Arnowitt-Deser-Misner (ADM) Hamiltonian formulation of general relativity. The phase space variables for gravity and matter are   $(q_{ab}, \pi^{ab})$ and $(\phi, p_\phi)$ respectively. The canonical action is   
\be
 S= \int dt\int d^3x \left( \pi^{ab} \dot{q}_{ab}  + p_\phi\dot{\phi} - N H - N^a C_a \right),
\ee
where $H$ and $C_a$ are the Hamiltonian and diffeomorphism constraints. The first is quadratic in all momentum  and the second is linear. Evolution is specified  by an arbitrary choice of lapse function $N$, which indicates the arbitrariness of the time coordinate.  

Let us now make two observations about this general setting. The first is that  any idea of ``evolving constants"  must also make reference to the arbitrary lapse function, and second, that making a time choice is equivalent to selecting a function on the phase space to represent time.   The latter identifies  a physical Hamiltonian as the (negative of the) phase space function conjugate to the chosen time function. Identifying this Hamiltonian requires solving the Hamiltonian constraint.  The key point is that the Hamiltonian so obtained is in general an explicit function of time if gravitational variables are used as clocks (and in most cases also for matter clocks). This applies not just to this action, but to any canonical action, including the ones used to model the evolution of the fine structure constant  \cite{Sandvik:2001rv}.

 This is readily illustrated in the cosmological setting where the gravitational phase space variables are the scale factor and its conjugate momentum $(a,P_a)$. The symmetry reduced canonical action for FLRW cosmology with a scalar field is 
 \be
S =  \int dt \Big( P_a \dot{a} + \PP \dot{\phi} - N\H \Big), \label{FRW}
\ee
where the Hamiltonian constraint is 
\be
\label{hcfrw}
\H =  -\dfrac{P_a^2}{24 a }  + a^3 \Lambda    + \dfrac{\PP^2}{2 a^3} + a^3 V(\phi) \approx 0.
\ee
 
 In the expansion phase of the Universe, a useful clock is volume time defined by   $t =  a^3$. Preserving this condition under time evolution fixes the lapse $N$:
 \be
 \dot{t} = 1 = \{ a^3, N\H\} \ \ \  \implies  N= \{ a^3, \H\}^{-1} , 
 \ee
 which gives $\displaystyle N=-\frac{4}{aP_a}$. Now since $a^3$ is time, we solve the Hamiltonian constraint for $P_a$, and substitute the result, and $a^3=t$, back into the action (\ref{FRW}).  The result is 
 \be
 S = \int dt \Big(  \PP \dot{\phi} - H_p \Big), \label{g-act}
 \ee
 where  
  \be
\label{sqrt_Ham}
H_p =  \sqrt{\dfrac{8}{3}} \sqrt{\Lambda  + \dfrac{\PP^2}{2 t^2} + V(\phi)};
\ee
 we chose the positive sign in the solution for $P_a$ when solving the Hamiltonian constraint. In this process (which is standard for gauge fixing of constrained systems) the gravity variables $(a,P_a)$ have been eliminated by choosing volume time and solving the constraint. 
 
 \section{Evolving constants} 
 
Let us now turn to the central question we would like to address: What is the relation of coupling constants of matter  as we observe them, to the constants appearing in the potential in the above gauge fixed action? We define matter couplings constants, such as charges and particle masses,  through local experiments. These experiments are typically localized in both space and time. Observations from  experiments are then compared with theories that are  Poincar\'e invariant within the localization defined by the experiments. This localization is not microscopic; Poincar\'e invariant field theories are believed to be applicable over large patches of the Universe, with distances extending to galactic size, and time scales to many years, at least in the current epoch of the Universe. For this reason, Poincar\'e invariance has to be relevant for at least some extended times and regions over which typical experiments are performed, and should be an emergent  feature of any complete theory describing the dynamics of gravity and  matter. 

How is the time gauge fixed action (\ref{g-act}) to be compared with a Poincar\'e covariant theory in a patch of the Universe (which does not have a square root Hamiltonian), and how might this lead to evolving constants? The answer to the first problem involves dealing with  the square root Hamiltonian (\ref{sqrt_Ham}). Fortunately, this can be done by noting that the currently observed ratio of matter to $\Lambda$ content of the Universe is $\Omega_M/\Omega_\Lambda \approx 0.45$ \cite{Ade:2015xua}. Therefore we can factor out $\Lambda$ and expand the square root to get
\be
\label{Hexp1}
\H_p =  \sqrt{\dfrac{8 \Lambda}{3}}  +\sqrt{\dfrac{2}{3\Lambda}} \Bigg[ \dfrac{\PP^2}{2 t^2} + \dfrac{1}{2}  m^2 \phi^2 + \lambda \phi^4 \Bigg] +  {\cal O} \left(\frac{\Omega_M}{\Omega_\Lambda}\right)^2,
\ee
where for concreteness we have fixed the potential $V(\phi)$ to be the mass term with a $\phi^4$ self-interaction.  The leading term is still time dependent through the kinetic term, and so a direct comparison with low energy late time physics is so far still not possible.

To see how to deal with this last issue, let us consider the canonical transformation  \cite{Hassan:2018psq}
\be
\tphi = t \phi ~,~~ \tPP = \dfrac{\PP}{t}.
\ee
Under this  change\footnote{This transformation is generated by the generating function $F_2(\phi, \tPP, t) = \phi \tPP t$.} the Lagrangian becomes
\bea
\PP \dot{\phi} - \H_p(\phi,\PP; t) &=& \tPP \dot{\tphi} -  \left( \dfrac{\tPP^2}{2} + \dfrac{1}{2}  \tm^2 \tphi^2 + \tlambda \tphi^4 + \dfrac{\tphi \tPP}{t}\right), \label{new_Ham}
\eea
where we have defined the ``low-energy" (large Universe) coupling constants 
\be
\tm(t) = \dfrac{m}{t} ~,~~ \tlambda(t) = \dfrac{\lambda}{t^4}. \label{couplingscale}
\ee
Now for $t\gg 1$ the last term in (\ref{new_Ham}) may be neglected and the result is the standard Hamiltonian for the  scalar field in flat 3-space. For comparison  the general expression is 
 \be
H_\phi = \int d^3x \left(\frac{p_\phi^2}{2\sqrt{q}} + \sqrt{q} q^{ab} \partial_a\phi\partial_b\phi + \sqrt{q}V(\phi)  \right), 
\ee
and the Hamiltonian (\ref{new_Ham}) also arises from this (for $t\gg1$)  by setting the spatial metric $q_{ab}$ to be the flat Euclidean metric, and restricting to the homogeneous field $\phi = \phi(t)$ appropriate for FLRW cosmology. The only new feature is the time dependence of the matter coupling constants given by (\ref{couplingscale}).  

This result demonstrates explicitly how time dependent matter couplings arise from a global choice of time, at least in the cosmological setting, and also how an emergent flat space Hamiltonian emerges.

\section{Discussion}

We described a method for obtaining a matter theory with evolving coupling constants from the standard Einstein-scalar theory action. This involved a global choice of time, an expansion of a square root Hamiltonian, and a canonical transformation. Unlike other approaches, no new fields,  couplings or actions were necessary. The result provides a proof of concept of the basic idea we propose.  

A similar method may be followed for the gravity-electromagnetic-Dirac theory, at least in the cosmological context. This would provide a derivation of an evolving fine structure constant through variation of the electric charge. The method is of course applicable  for any matter coupled to general relativity (e.g., if a Higgs field is added, this method would lead to a time-dependent Higgs mass and vacuum expectation value), and may be potentially useful for the inhomogeneous case as well, at least where the latter are treated as perturbations.

The exact time dependence of the scalings (\ref{couplingscale}) depends on the time gauge chosen. In a different time gauge, this scaling would be different. For the volume time gauge we use here, both the mass and quartic coupling constant decrease as the Universe evolves. It may turn out that some clocks work better than others in describing the physics that we see. This is not necessarily a problem of gauge dependence -- the CMB exhibits large scale homogeneity and isotropy only in one particular reference frame from an infinite number of choices. Similarly, if experimental evidence does end up supporting the evolution of  matter coupling constants, volume time, and perhaps this scheme altogether, may be disproved if the observed variations do not correspond to the computed variations.

Lastly we note that in the method we have described, evolving coupling constants arise naturally, and indeed are unavoidable, because global time gauges generically lead to time dependent physical Hamiltonians. The only exception to this rule at present appears to be dust time \cite{Husain:2011tk}, where the corresponding physical Hamiltonian is time-independent.
 
\vspace{1cm}

\noindent{\bf Acknowledgments} B.Q is supported by the Fulbright fellowship. S.M.H was supported by the Lewis Doctoral Fellowship.

\bibliography{ChangingConstants}

\end{document}